GEOCHEMISTRY

# Gallium isotopic evidence for extensive volatile loss from the Moon during its formation

Chizu Kato[1]*[†] and Frédéric Moynier[1,2]

The distribution and isotopic composition of volatile elements in planetary materials holds a key to the characterization of the early solar system and the Moon's formation. The Moon and Earth are chemically and isotopically very similar. However, the Moon is highly depleted in volatile elements and the origin of this depletion is still debated. We present gallium isotopic and elemental measurements in a large set of lunar samples to constrain the origin of this volatile depletion. We show that while Ga has a geochemical behavior different from zinc, both elements show a systematic enrichment in the heavier isotopes in lunar mare basalts and Mg-suite rocks compared to the silicate Earth, pointing to a global-scale depletion event. On the other hand, the ferroan anorthosites are isotopically heterogeneous, suggesting a secondary distribution of Ga at the surface of the Moon by volatilization and condensation. The isotopic difference of Ga between Earth and the Moon and the isotopic heterogeneity of the crustal ferroan anorthosites suggest that the volatile depletion occurred following the giant impact and during the lunar magma ocean phase. These results point toward a Moon that has lost its volatile elements during a whole-scale evaporation event and that is now relatively dry compared to Earth.

## INTRODUCTION

The Moon is thought to have formed from an impact between the proto-Earth and a Mars-sized impactor (1–3) or from multiple impactors (4), which resulted in a formation of a silicate vapor–rich disk (5) that later condensed to form the Moon (6). Most lunar rocks are depleted in volatile elements compared to the bulk silicate Earth (BSE) (7–9), pointing toward a volatile-depleted lunar interior, although the origin of the depletion is poorly understood (10). This general volatile depletion is more puzzling, considering that not all samples point toward a volatile-depleted Moon. For example, some olivine from the lunar pyroclastic glasses contains melt inclusions with elevated abundances of OH, F, and Cl, implying a source for these samples enriched in volatiles (11, 12). In addition, some mare basalt apatites have high OH content (13–15). However, recent measurements on plagioclase feldspar, which is a major mineral of the lunar crust, indicated that much of the lunar mantle must have <100 parts per million (ppm) of water (16).

Recent high-precision isotopic measurements of lunar samples tried to solve this problem by using Zn, Cl, Rb, and K isotopes as tracers of the origins of the volatile depletion in mare basalts (17–24). These studies have demonstrated that mare basalts are depleted in the lighter isotopes of those elements compared to the BSE, suggesting a whole-scale evaporation event such as the giant impact (20, 24), and/or during magma ocean differentiation (10, 21, 22), or during an eruptive degassing of volatiles from mare basalt lavas into a vacuum (19).

To further investigate the volatile elements in the Moon and their depletion, we analyzed, with high precision, Ga isotopic and elemental abundances on various lunar samples that were previously analyzed for their Zn isotopic composition (21). Gallium is a moderately volatile element with a 50% condensation temperature of 968 K (25), which shows a ~10-fold depletion in lunar basalts compared to their terrestrial counterparts (7). Previous studies of different crustal and magmatic terrestrial samples demonstrated that Ga isotopes do not fractionate during magmatic processes within analytical uncertainty (26). The relatively low condensation temperature and the absence of isotopic fractionation during magmatic processes make the Ga isotope ratio a powerful tracer of high-temperature evaporation during volatile depletion events on planets.

## RESULTS

Gallium concentrations and isotopic compositions ($\delta^{71}$Ga = ([$^{71}$Ga/$^{69}$Ga]$_{sample}$/[$^{71}$Ga/$^{69}$Ga]$_{Ga\ IPGP\ standard}$ − 1) × 1000) are reported in Table 1 and Fig. 1. There is a moderately good correlation between the Ga and the Zn isotopic composition of the lunar samples (Fig. 2). Because Zn (lithophile) and Ga (siderophile) do not have similar geochemical behaviors during planetary differentiation or igneous processes (Ga is incompatible, whereas Zn is fairly compatible), the correlation between $\delta^{66}$Zn and $\delta^{71}$Ga must be a consequence of volatile-related processes. This observation is a strong argument in favor of a volatile control on both the abundance and isotopic composition of Ga and Zn. The Ga concentrations of the different lunar samples are all below 4 ppm. Four high-Ti basalts ($\delta^{71}$Ga = 0.15 to 0.57‰) and three low-Ti basalts ($\delta^{71}$Ga = 0.09 to 0.32‰) demonstrate that the mare basalts are enriched in the heavier isotopes of Ga compared to the BSE [$\delta^{71}$Ga = 0.00 ± 0.06‰ 2 SD; (26)] (Fig. 1). A few exceptions (two of nine samples) are isotopically light, including 14053, which coincidently also has one of the lightest Zn isotopic compositions (21). These Zn-rich samples cannot represent volatile-rich, undegassed regions of the lunar mantle because there would be no reason for these samples to be isotopically light. On the other hand, this enrichment in the lighter isotopes of Ga and Zn is best explained by contamination of the samples from the condensation of an isotopically light vapor at the surface of the samples (21) and has therefore been excluded from further discussion. Gallium concentrations in the mare basalts span a small range from 0.8 to 3.2 ppm, which is more than seven times less than their terrestrial counterparts [~20 ppm; (26)]. Compared to the mare basalts, ferroan anorthosites (FANs) display larger Ga isotopic variability ($\delta^{71}$Ga between −0.47 and 0.85‰; Fig. 1). In comparison, the Mg-suite rock (77215) was isotopically heavy ($\delta^{71}$Ga = 0.16 ± 0.05‰ 2 SD). The pyroclastic green

[1]Institut de Physique du Globe de Paris, Université Paris Diderot, Institut Universitaire de France, Paris, France. [2]Institut Universitaire de France, 75005 Paris, France.
*Present address: Division of Sustainable Energy and Environmental Engineering, Graduate School of Engineering, Osaka University, Suita, Osaka 565-0871, Japan.
†Corresponding author. Email: kato@ipgp.fr

Table 1. **Gallium isotopic composition and element concentration of the lunar samples.** $n$ = number of isotopic measurements.

| Sample type | Sample no. | $\delta^{71}$Ga (‰) | 2 SD | $n$ | Ga (ppm) | Ga (ppm)* |
|---|---|---|---|---|---|---|
| Mg-suite | | | | | | |
| Cataclastic norite | 77215 | 0.16 | 0.05† | 1 | 1.5 | — |
| Regolith | | | | | | |
| Soil | 78221 | 1.16 | 0.05† | 1 | 2.1 | — |
| Pyroclastic glass | | | | | | |
| Green glass | 15426 | −0.35 | 0.05† | 1 | 1.3 | 4.7 |
| FANs | | | | | | |
| Anorthosite (with melt) | 62255 | −0.47 | 0.05† | 1 | 0.5 | — |
| Cataclastic anorthosite | 60015 | 0.17 | 0.05† | 1 | 1.4 | — |
| Ferroan anorthosite | 15415 | 0.85 | 0.05† | 1 | 1.1 | 3.1 |
| Noritic anorthosite | 67955 | 0.20 | 0.05† | 1 | 1.9 | 3.5–4.2 |
| Low-Ti basalts | | | | | | |
| Basalt (low Ti) | 15499 | 0.09 | 0.05 | 2 | 3.2 | 3 |
| Olivine basalt | 12012 | 0.32 | 0.02 | 2 | 1.1 | — |
| Olivine basalt | 12040 | 0.15 | 0.05 | 3 | 1.4 | 1.9 |
| Mare basalt (Al-rich) | 14053 | −0.17 | 0.05† | 1 | 0.8 | 2.5–4.8 |
| High-Ti basalts | | | | | | |
| Basalt (high Ti) | 70017 | 0.15 | 0.01 | 3 | 2.5 | 3.1–21 |
| Basalt (high Ti) | 10003 | 0.35 | 0.06 | 2 | 1.7 | 4.0–4.7 |
| Basalt (high Ti) | 70135 | 0.57 | 0.02 | 2 | 2.5 | 7–16 |
| Ilmenite basalt (high K) | 10057 | −0.03 | 0.03 | 6 | — | 4.8 |
| Ilmenite basalt (high Mg/Fe) | 12005 | 0.22 | 0.01 | 3 | 1.1 | — |
| BSE | | 0.00‡ | 0.05 | | 3.9§ | — |

*Data from Meyer (47).  †Used the 2 SD estimated from Kato et al. (26).  ‡Data from Kato et al. (26).  §Data from McDonough (48).

glass (15426) is isotopically lighter than the mare basalts ($\delta^{71}$Ga = −0.35 ± 0.05‰ 2 SD). Finally, one regolith sample (78221) had an isotopic value of $\delta^{71}$Ga = 1.16 ± 0.05‰ (2 SD), being the heaviest sample within this sample set. This enrichment in the heavy isotope of Ga in a regolith sample was expected on the basis of previous Zn (17), S (27), and K (28) isotope studies that show that the lighter isotopes were lost during the regolith gardening by either sputtering or micrometeorite impacts.

## DISCUSSION

The source of the mare basalts is enriched in the heavy isotope of Ga ($\delta^{71}$Ga = 0.09 to 0.57‰, $n$ = 7) compared to the BSE [$\delta^{71}$Ga = 0.00 ± 0.06‰ 2 SD; (26)]. Together with previous results based on Zn [$\delta^{66}$Zn = 1.4 ± 0.5‰; (21)] and K [only two data in mare basalts: $\delta^{41}$K = 0.372 ± 0.061‰ and $\delta^{41}$K = 0.460 ± 0.037‰; (24)], this confirms the general enrichment of mare basalts in heavy isotopes of moderately volatile elements compared to Earth. It has previously been suggested that the heavy isotope enrichments in the mare basalts compared to the BSE in Zn were due to the fact that the source of the mare basalts was isotopically depleted by a global-scale high-temperature evaporation event (20). Isotopic enrichment and concentration depletion compared to the BSE is seen in the mare basalts for Ga, along with a variety of elements (Zn, K, S, Cl, and Rb), regardless of different sampling sites, further confirms a global-scale volatile depletion event, such as the giant impact or during evaporation from the magma ocean as the cause.

The Mg-suite sample consists of igneous cumulates that are characterized by their high Mg# together with anorthitic plagioclase, low abundance of compatible elements (for example, Ni and Co), and a small K, rare-earth elements and P (KREEP) component (29). Although the origin of the Mg-suite is debated, their plutonic origin implies that the isotopic fractionation of Ga did not occur via evaporation during degassing into a vacuum. The Ga isotopic composition of the Mg-suite sample is similar to the mare basalts and shows a slight enrichment in the heavy isotope compared to the BSE. This further confirms that the source magma was already depleted in the volatile elements, and therefore,

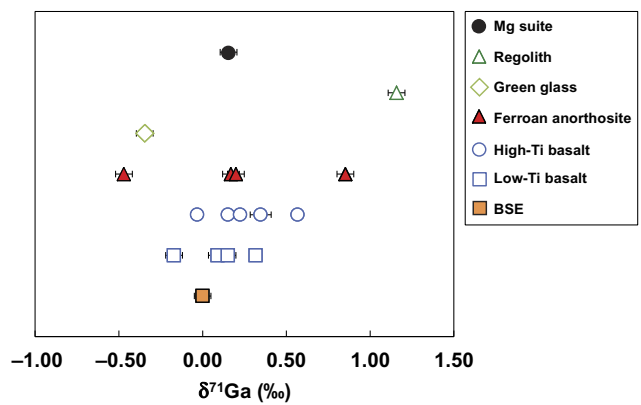

**Fig. 1. Gallium isotopic compositions for the lunar samples.** BSE value from Kato *et al.* (*26*).

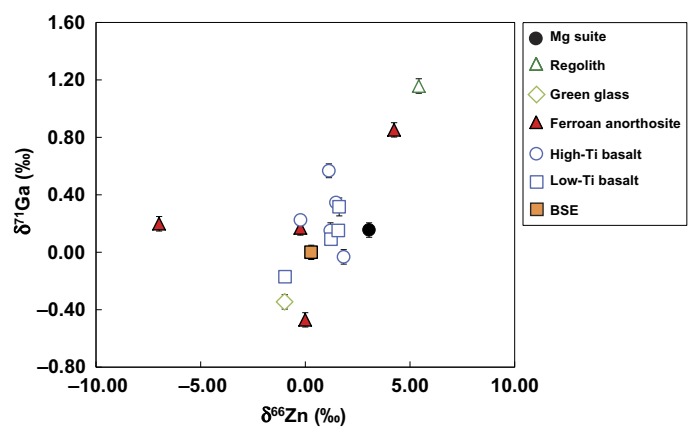

**Fig. 2. Gallium versus Zn isotopic composition of the lunar samples.** BSE values are from Kato *et al.* (*26*) (Ga) and Chen *et al.* (*46*) (Zn). Lunar Zn isotopic values are from Kato *et al.* (*21*).

**Table 2. Ga/Al ratio of the FANs.**

| Sample type | Sample no. | Ga (ppm) | Al (ppm)* | Ga/Al*$10^4$ (ppm/ppm) |
|---|---|---|---|---|
| FANs | | | | |
| Anorthosite (with melt) | 62255 | 0.5 | 160786 | 0.03 |
| Cataclastic anorthosite | 60015 | 1.4 | 185776 | 0.08 |
| Ferroan anorthosite | 15415 | 1.1 | 188440 | 0.06 |
| Noritic anorthosite | 67955 | 1.9 | 143674 | 0.13 |

*Data from Meyer (*47*).

the heavy Ga, Zn, K, and Cl isotopic compositions of mare basalts are not due to degassing of their lava in a vacuum.

The lunar pyroclastic green glass (15426) is highly enriched in volatile elements (including Ga and Zn) at the surface compared to the interior (*30–32*), which is also enriched in the lighter isotopes of Ga and Zn ($\delta^{71}Ga = -0.35‰$, $\delta^{66}Zn = -0.98‰$). These results further confirm that they were formed from lunar fire fountaining from the lunar interior (*33*, *34*). During the ascent of the magma, the pressure decrease induces the formation of gas bubbles enriched in the lighter isotopes of Ga and Zn due to kinetic fractionation. This volatile-rich and isotopically light gas then condenses onto the surface of the quenched glass after eruption.

FANs are considered to have been produced from the crystallization of the lunar magma ocean as flotation lunar crust (*35*). They would therefore represent the primary crust of the Moon and are the oldest rock samples of the Moon (*36*). The isotopic variation in a single sample set is the largest seen within the samples analyzed in this study, with a maximum range of 1.3‰ between the brecciated 62255 ($\delta^{71}Ga = -0.47‰$) and unbrecciated pristine 15415 ($\delta^{71}Ga = 0.85‰$). Because Ga isotopes do not fractionate (within analytical uncertainty) during fractional crystallization (*26*), two scenarios can account for the Ga isotopic variability recorded in FANs: (i) the isotopic variations are inherited from the heterogeneity of the original source magma, or (ii) secondary redistribution of Ga at the surface of the Moon fractionated Ga isotopes. Gallium is moderately incompatible and behaves similarly to Al during magmatic differentiation due to their comparable ionic radii, valence, and ionization potential, and therefore, the Ga/Al ratios do not fractionate during the last stages of crystallization or by partial melting (*37*). The FANs exhibit variable Ga/Al ratios (Table 2), which suggests that the difference in the Ga/Al ratio is rather due to regional volatilization and condensation. Impact on the surface of the FANs will selectively evaporate the light isotope, and later, the isotopically light gas will condense onto another surface, forming the isotopically heterogeneous FANs. In addition, some FANs may represent mixed lithologies (*38*), and therefore, it is possible that the observed trend represents secondary mixing between lithologies with variable isotopic compositions. Zinc and Ga isotope ratios follow systematic variations between FAN samples (Fig. 2), further suggesting that high-temperature evaporation events lead to the origin of these variations (*21*). In addition, the Fe/Zn ratio, which also does not fractionate during partial melting under reducing conditions (*39*), can be used as a tracer of volatile depletion because Zn is more volatile than Fe (*25*). Following this logic, the low Fe/Zn ratio of lunar mare basalts compared to terrestrial basalts was used as an evidence for the volatile depletion of the Moon (*40*). Similarly, the variations between FAN samples analyzed here suggest volatilization as the cause for the isotopic fractionation. This interpretation is confirmed by the texture of the samples; the unbrecciated pristine 15415 (*41*) is enriched in heavy isotopes for both Ga ($\delta^{71}Ga = 0.85‰$) and Zn ($\delta^{66}Zn = 4.24‰$), whereas the brecciated 62255 ($\delta^{71}Ga = -0.47‰$; $\delta^{66}Zn = 0.00‰$) incorporated the isotopically light Ga- and Zn-rich vapor during cataclasis or exposure on the lunar surface. This concludes that the FANs do not represent the isotopic composition of their source magma, and this composition is controlled by billions of years of impacts on the surface of the Moon.

Mare basalts and the plutonic Mg-suite sample are enriched in the heavy isotopes of Ga and Zn compared to the BSE. This supports a magma source originally depleted in volatile elements and not volatile loss during degassing into a vacuum. Because all chondrites are isotopically lighter in Ga isotopes than the BSE (and therefore lighter than the Moon) (*42*), this heavy isotope enrichment cannot be due to the composition of the impactor. In turn, it would require a whole-scale evaporation event for the loss of Ga, either by the giant impact or during the lunar magma ocean. Outgassing during planetary impacts and/or magma ocean phases plays a crucial role in setting the volatile inventories of planets. For the Earth-Moon system, it has been shown that hydrodynamic escape from the lunar disk is not possible and that volatiles (including water) cannot be lost to space (43). On the other hand, dynamical biphased vapor-melt numerical modeling of the proto-lunar disk shows that because of magnetorotational instability, the vapor phase is viscous and accretes preferentially to Earth, leaving volatile-depleted material to form the Moon (6). Furthermore, the volatile loss during a magma ocean phase of the Moon would predict that this effect would be even more pronounced on low mass bodies. This is observed in the angrite and HED (howardite-eucrite-diogenite) parent bodies, which are both highly depleted in volatile elements. Therefore, if the volatile loss occurred during the magma ocean phase, it would be expected that Ga isotopes would be highly fractionated in HED and angrite meteorites. Regardless of which whole-scale evaporation event occurred, our results imply that planets accreted principally from highly differentiated planetesimals and must have accreted volatile depleted.

## MATERIALS AND METHODS

Gallium isotope compositions were measured in four low-Ti mare basalts (12012, 12040, 14053, and 15499), five high-Ti mare basalts (10003, 10057, 12005, 70017, and 70135), one Mg-suite plutonic rock (77215), one regolith sample (78221), one pyroclastic green glass (15426), and four FANs (15415, 60015, 62255, and 67955). All these samples have been previously studied for their Zn isotopic composition (21). The mare basalt samples were collected during the Apollo 11, 12, 14, 15, and 17 missions, therefore covering a wide area of the Moon. Pyroclastic green glasses are unusual lunar samples that are highly enriched in volatile elements at their surface, which is usually interpreted as formed in a lava fountain and recovered by the condensation of an isotopically light vapor following their formation. FANs are the first lunar flotation cumulate crust that crystallized from the lunar magma ocean (44). Magnesium-suite plutonic rocks are intrusive igneous rocks that have higher Mg/Fe ratios compared to the FAN crust (45).

Up to 1-g chips of samples were powdered with an agate mortar and pestle. Samples were dissolved in a mixture of HF and $HNO_3$ (3:1) for 3 days in Teflon beakers at 120°C. The Ga and matrix cut were collected from the Zn purification chemistry and were dried down and dissolved in 1 ml of 6 mol HCl for gallium purification that was achieved following the method of Kato et al. (26). The samples were analyzed for Ga concentration and isotopic compositions using a Thermo Fisher Scientific Neptune Plus multicollector inductively coupled plasma mass spectrometer (MC-ICPMS) at the Institut de Physique du Globe de Paris (IPGP). Gallium isotopic variations were shown in permil deviations from the Ga IPGP standard, which was used here instead of the international standard (National Institute of Standards and Technology Standard Reference Material 3119a) that is fractionated by 1.4‰ compared to major terrestrial samples [see discussion of Kato et al. (26)]. Samples were bracketed with the Ga IPGP standard. The analytical precision on the $\delta^{71}Ga$ was estimated as the 2 SD from the replicated measurements for each sample. The 2 SD estimated by Kato et al. (26) was used for samples for which it was only possible to perform one measurement. The Ga concentrations were obtained by comparing the intensity of the sample with that of a known concentration, with an estimated precision of ±10%. The total blank of the procedure was <0.02 ng, which is negligible compared to the amount of Ga in each sample (>25 ng).

## REFERENCES AND NOTES

1. W. K. Hartmann, D. R. Davis, Satellite-sized planetesimals and lunar origin. *Icarus* **24**, 504–515 (1975).
2. R. M. Canup, Forming a Moon with an Earth-like composition via a giant impact. *Science* **338**, 1052–1055 (2012).
3. M. Ćuk, S. T. Stewart, Making the Moon from a fast-spinning Earth: A giant impact followed by resonant despinning. *Science* **338**, 1047–1052 (2012).
4. R. Rufu, O. Aharonson, H. B. Perets, A multiple-impact origin for the Moon. *Nat. Geosci.* **10**, 89–94 (2017).
5. C. Visscher, B. Fegley Jr., Chemistry of impact-generated silicate melt-vapor debris disks. *Astrophys. J. Lett.* **767**, L12 (2013).
6. S. Charnoz, C. Michaut, Evolution of the protolunar disk: Dynamics, cooling timescale and implantation of volatiles onto the Earth. *Icarus* **260**, 440–463 (2015).
7. P. H. Warren, The Moon, in *Meteorites, Comets, and Planets Vol. 1*, A. M. Davis Ed. (Elsevier-Pergamon, 2003), pp. 559–599.
8. S. R. Taylor, G. J. Taylor, L. A. Taylor, The Moon: A Taylor perspective. *Geochim. Cosmochim. Acta* **70**, 5904–5918 (2006).
9. R. Wolf, E. Anders, Moon and Earth: Compositional differences inferred from siderophiles, volatiles, and alkalis in basalts. *Geochim. Cosmochim. Acta* **44**, 2111–2124 (1980).
10. J. M. D. Day, F. Moynier, Evaporative fractionation of volatile stable isotopes and their bearing on the origin of the Moon. *Philos. Trans. R. Soc. A* **372**, 20130259 (2014).
11. A. E. Saal, E. H. Hauri, J. A. Van Orman, M. J. Rutherford, Hydrogen isotopes in lunar volcanic glasses and melt inclusions reveal a carbonaceous chondrite heritage. *Science* **340**, 1317–1320 (2013).
12. E. H. Hauri, A. E. Saal, M. J. Rutherford, J. A. Van Orman, Water in the Moon's interior: Truth and consequences. *Earth Planet. Sci. Lett.* **409**, 252–264 (2015).
13. F. M. McCubbin, A. Steele, E. H. Hauri, H. Nekvasil, S. Yamashita, R. J. Hemley, Nominally hydrous magmatism on the Moon. *Proc. Natl. Acad. Sci. U.S.A.* **107**, 11223–11228 (2010).
14. J. W. Boyce, Y. Liu, G. R. Rossman, Y. Guan, J. M. Eiler, E. M. Stolper, L. A. Taylor, Lunar apatite with terrestrial volatile abundances. *Nature* **466**, 466–469 (2010).
15. R. Tartèse, M. Anand, Late delivery of chondritic hydrogen into the lunar mantle: Insights from mare basalts. *Earth Planet. Sci. Lett.* **361**, 480–486 (2013).
16. R. D. Mills, J. I. Simon, C. M. O. D. Alexander, J. Wang, E. H. Hauri, Water in alkali feldspar: The effect of rhyolite generation on the lunar hydrogen budget. *Geochem. Perspect. Lett.* **3**, 115–123 (2017).
17. F. Moynier, F. Albarède, G. F. Herzog, Isotopic composition of zinc, copper, and iron in lunar samples. *Geochim. Cosmochim. Acta* **70**, 6103–6117 (2006).
18. G. F. Herzog, F. Moynier, F. Albarède, A. A. Berezhnoy, Isotopic and elemental abundances of copper and zinc in lunar samples, Zagami, Pele's hairs, and a terrestrial basalt. *Geochim. Cosmochim. Acta* **73**, 5884–5904 (2009).
19. Z. D. Sharp, C. K. Shearer, K. D. McKeegan, J. D. Barnes, Y. Q. Wang, The chlorine isotope composition of the Moon and implications for an anhydrous mantle. *Science* **329**, 1050–1053 (2010).
20. R. C. Paniello, J. M. D. Day, F. Moynier, Zinc isotopic evidence for the origin of the Moon. *Nature* **490**, 376–379 (2012).
21. C. Kato, F. Moynier, M. C. Valdes, J. K. Dhaliwal, J. M. D. Day, Extensive volatile loss during formation and differentiation of the Moon. *Nat. Commun.* **6**, 7617 (2015).
22. J. W. Boyce, A. H. Treiman, Y. Guan, C. Ma, J. M. Eiler, J. Gross, J. P. Greenwood, E. M. Stolper, The chlorine isotope fingerprint of the lunar magma ocean. *Sci. Adv.* **1**, e1500380 (2015).
23. E. Pringle, F. Moynier, Rubidium isotopic composition of the Earth, meteorites, and the Moon: Evidence for the origin of volatile loss during planetary accretion. *Earth Planet. Sci. Lett.* **473**, 62–70 (2017).
24. K. Wang, S. B. Jacobsen, Potassium isotopic evidence for a high-energy giant impact origin of the Moon. *Nature* **538**, 487–490 (2016).
25. K. Lodders, Solar system abundances and condensation temperatures of the elements. *Astrophys. J.* **591**, 1220–1247 (2003).
26. C. Kato, F. Moynier, J. Foriel, F.-Z. Teng, I. S. Puchtel, The gallium isotopic composition of the bulk silicate Earth. *Chem. Geol.* **448**, 164–172 (2017).
27. H. G. Thode, C. E. Rees, Measurement of sulphur concentrations and the isotope ratios $^{33}S/^{32}S$, $^{34}S/^{32}S$, and $^{36}S/^{32}S$ in Apollo 12 samples. *Earth Planet. Sci. Lett.* **12**, 434–438 (1971).
28. M. Humayun, R. N. Clayton, Precise determination of the isotopic composition of potassium: Application to terrestrial rocks and lunar soils. *Geochim. Cosmochim. Acta* **59**, 2115–2130 (1995).
29. P. H. Warren, Anorthosite assimilation and the origin of the Mg/Fe-related bimodality of pristine moon rocks: Support for the magmasphere hypothesis. *J. Geophys. Res.* **91**, 331–343 (1986).
30. H. Wänke, H. Baddenhausen, G. Dreibus, E. Jagoutz, H. Kruse, H. Palme, B. Spettel, F. Teschke, Multielement analyses of Apollo 15, 16, and 17 samples and the bulk composition of the moon. *Geochim. Cosmochim. Acta* **2**, 1461–1481 (1973).

31. C. J. Meyer Jr., D. S. McKay, D. H. Anderson, P. J. Butler Jr., The source of sublimates on the Apollo 15 green and Apollo 17 orange glass samples. *Proc. Lunar Sci. Conf. 6th* **2**, 1673–1699 (1975).
32. J. T. Wasson, W. V. Boynton, G. W. Kallemeyn, L. L. Sundberg, C. M. Wai, Volatile compounds released during lunar lava fountaining. *Proc. Lunar Sci. Conf. 7th* **2**, 1583–1595 (1976).
33. W. I. Ridley, A. M. Reid, J. L. Warner, R. W. Brown, Apollo 15 green glasses. *Phys. Earth Planet. Inter.* **7**, 133–136 (1973).
34. R. A. Fogel, M. J. Rutherford, Magmatic volatiles in primitive lunar glasses: I. FTIR and EPMA analyses of Apollo 15 green and yellow glasses and revision of the volatile-assisted fire-fountain theory. *Geochim. Cosmochim. Acta* **59**, 201–215 (1995).
35. P. H. Warren, The magma ocean concept and lunar evolution. *Annu. Rev. Earth Planet. Sci.* **13**, 201–240 (1985).
36. L. E. Borg, J. N. Connelly, M. Boyet, R. W. Carlson, Chronological evidence that the Moon is either young or did not have a global magma ocean. *Nature* **477**, 70–72 (2011).
37. R. De Argollo, J. G. Schilling, Ge-Si and Ga-Al fractionation in Hawaiian volcanic rocks. *Geochim. Cosmochim. Acta* **42**, 623–630 (1978).
38. O. B. James, M. M. Lindstrom, J. J. McGee, Lunar ferroan anorthosite 60025: Petrology and chemistry of mafic lithologies. *Proc. Lunar Planet. Sci.* **21**, 63–87 (1991).
39. C.-T. A. Lee, P. Luffi, V. Le Roux, R. Dasgupta, F. Albarède, W. P. Leeman, The redox state of arc mantle using Zn/Fe systematics. *Nature* **468**, 681–685 (2010).
40. F. Albarède, E. Albalat, C.-T. A. Lee, An intrinsic volatility scale relevant to the Earth and Moon and the status of water in the Moon. *Meteorit. Planet. Sci.* **50**, 568–577 (2015).
41. R. Ganapathy, J. W. Morgan, U. Krähenbühl, E. Anders, Ancient meteoritic components in lunar highland rocks: Clues from trace elements in Apollo 15 and 16 samples. *Proc. Lunar Sci. Conf.* **4**, 1239–1261 (1973).
42. C. Kato, J. Foriel, F. Moynier, Isotopic study of gallium in terrestrial and meteorite samples. *LPI Contrib.* **1800**, 5209 (2014).
43. M. Nakajima, D. J. Stevenson, Hydrodynamic escape does not prevent the "wet" Moon formation. *Lunar Planet. Sci. Conf.* **45**, 2770 (2014).
44. C. K. Shearer, P. C. Hess, M. A. Wieczorek, M. E. Pritchard, E. M. Parmentier, L. E. Borg, J. Longhi, L. T. Elkins-Tanton, C. R. Neal, I. Antonenko, R. M. Canup, A. N. Halliday, T. L. Grove, B. H. Hager, D.-C. Lee, U. Wiechert, Thermal and magmatic evolution of the Moon. *Rev. Mineral. Geochem.* **60**, 365–518 (2006).
45. C. K. Shearer, J. J. Papike, Early crustal building processes on the moon: Models for the petrogenesis of the magnesian suite. *Geochim. Cosmochim. Acta* **69**, 3445–3461 (2005).
46. H. Chen, P. S. Savage, F.-Z. Teng, R. T. Helz, F. Moynier, Zinc isotope fractionation during magmatic differentiation and the isotopic composition of the bulk Earth. *Earth Planet. Sci. Lett.* **369–370**, 34–42 (2013).
47. C. Meyer, Lunar Sample Compendium (2012); www-curator.jsc.nasa.gov/lunar/lsc/index.cfm.
48. W. F. McDonough, Comment on "Abundance and distribution of gallium in some spinel and garnet lherzolites" by D. B. McKay and R. H. Mitchell. *Geochim. Cosmochim. Acta* **54**, 471–473 (1990).

**Acknowledgments:** We thank two anonymous reviewers for greatly improving the quality of this manuscript and to the editor, C.-T. Lee, for efficient edition. We are grateful to Curation and Analysis Planning Team for Extraterrestrial Materials (CAPTEM) and the NASA Johnson Space Center curatorial staff for provision of samples. We would like to thank N. Badullovich and J. B. Creech for proofreading the manuscript. **Funding:** F.M. acknowledges funding from the European Research Council (ERC) under the H2020 framework program/ERC grant agreement #637503 (Pristine), the financial support of the Labex UnivEarthS program at Sorbonne Paris Cité (ANR-10-LABX-0023 and ANR-11-IDEX-0005-02), and the ANR through a chaire d'excellence Sorbonne Paris Cité. Parts of this work were supported by IPGP multidisciplinary program PARI and by Region Île-de-France SESAME grant no. 12015908. **Author contributions:** C.K. collected the data. F.M. and C.K. conceived the project and wrote the manuscript. **Competing interests:** The authors declare that they have no competing interests. **Data and materials availability:** All data needed to evaluate the conclusions in the paper are present in the paper. Additional data related to this paper may be requested from the authors.

Submitted 22 February 2017
Accepted 27 June 2017
Published 28 July 2017
10.1126/sciadv.1700571

Citation: C. Kato, F. Moynier, Gallium isotopic evidence for extensive volatile loss from the Moon during its formation. *Sci. Adv.* **3**, e1700571 (2017).

# Science Advances

**Gallium isotopic evidence for extensive volatile loss from the Moon during its formation**
Chizu Kato and Frédéric Moynier